**Dataset in Focus**

# Dataset of Philippine Presidents' Speeches from 1935 to 2016

John Paul P. Miranda
Mexico Campus, Don Honorio Ventura State University, Philippines
jppmiranda@dhvsu.edu.ph
(corresponding author)

## Abstract

*Purpose* – The dataset was collected to examine and identify possible key topics within these texts.

*Method* – Data preparation such as data cleaning, transformation, tokenization, removal of stop words from both English and Filipino, and word stemming was employed in the dataset before feeding it to sentiment analysis and the LDA model.

*Results* – The topmost occurring word within the dataset is "development" and there are three (3) likely topics from the speeches of Philippine presidents: economic development, enhancement of public services, and addressing challenges.

*Conclusion* – The dataset was able to provide valuable insights contained among official documents. While the study showed that presidents have used their annual address to express their visions for the country. It also presented that the presidents from 1935 to 2016 faced the same problems during their term.

*Recommendations* – Future researchers may collect other speeches made by presidents during their term; combine them to the dataset used in this study to further investigate these important texts by subjecting them to the same methodology used in this study. The dataset may be requested from the authors and it is recommended for further analysis. For example, determine how the speeches of the president reflect the preamble or foundations of the Philippine constitution.

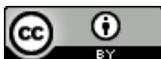

*Keywords* – data mining, Philippines, president, speech, text mining.

**Keywords:** dataset, Philippine president, speeches, data mining, quantitative text analysis, text mining, sentiment analysis, topic modeling

## Background of Dataset

There are 77 annual presidents' speeches delivered to the nation from 1935 to 2016. These were extracted from multiple sources: seventy-two (72) came from official government websites (i.e., Official Gazette of the Philippines and Presidential Communication Operations Office (PCOO)) and five (5) of them came from online news sources (i.e., ABS-CBN News, GMA News Online, Rappler). All of these speeches are in a text-based format. The dataset contains pre-cleaned and tokenized 656,873 words.

The study of Miranda and Bringula (2021) to which the dataset was fully utilized aims to analyze what possible themes or likely topics can be found from the presidents' speeches and how do these speeches by sentiment can change over time. The study sought to answer these problems by implementing a topic model and sentiment analysis to the dataset. LDA was previously used for analyzing personality traits of social media user (Liu, Wang, & Jiang, 2016), identifying and summarization of topics on large collection of documents (Blei, 2012; Rosen-Zvi et al., 2004), classifying key themes from comments on video tutorials (Miranda and Martin, 2020), and extraction of linguistic metaphor (Heintz et al., 2013). While several studies have used sentiment analysis for analyzing important topics on political issues (Stieglitz and Dang, 2012), understanding politicians' behavior towards social media discourse (DiGrazia et al., 2013), examining public mood towards events (Bollen et al., 2011), and determining the sentiments of YouTube learners through their feedback (Miranda & Martin, 2020; Bringula et al., 2019).

Figure 1 shows the entire data cleaning and preparation process implemented towards the 77 speeches. It involves six (6) main parts: combining all speeches, tokenizing the combined speeches, transforming them to lowercase letters, removing numbers and special characters (i.e., "*", "#"), applying word stemming, filtering of English and Filipino stop words from the Natural Language Toolkit (NLTK) library. The process resulted in the cleaned dataset which was also the corpus which consists of 656,873 words. A copy of the corpus was further subdivided per president (i.e., Arroyo: 2001-2009), this was implemented to determine the probable topics contained within presidents' speeches. The above-mentioned process was strictly followed before subjecting the corpus to sentiment analysis and topic model. For the sentiment analysis, the study used the sentiment intensity analyzer from the NLTK library to acquire and aggregate the polarity scores (i.e., positivity and negativity of sentences) and to see the changes in the sentiment per president. For the topic modeling, the dataset was subjected to multidimensional scaling to visualize how many topics can be inferred based on their inter-topic distances. The authors then decided after careful deliberation on the probable labels.

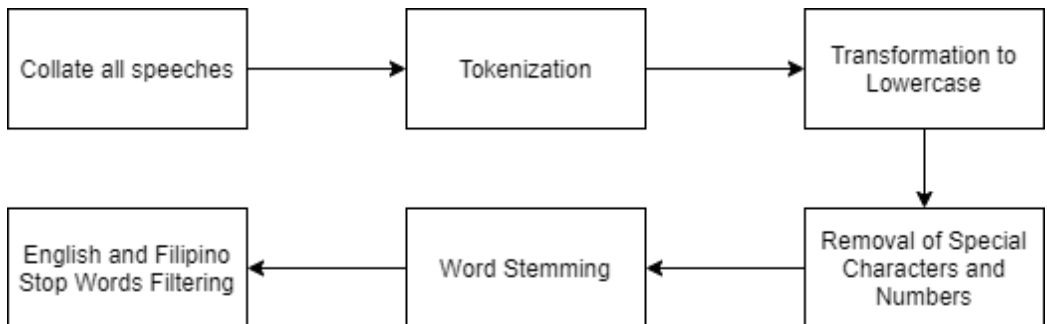

*Figure 1.* Step by step process of data preparation implemented to the speeches

## Results of Data Analysis

The researchers have used NLTK's Valence Aware Dictionary and Sentiment Reasoner (VADER) to perform the sentiment analysis in the dataset. VADER sentiment analysis is known for its simplicity and effectiveness (Hutto & Gilbert, 2014). While Latent Dirichlet Allocation (LDA) model was used to discover topic clusters from the dataset. LDA is recognized for its usefulness and popularity as a technique for topic modeling (Jelodar et al., 2019). Both models implement bag-of-words. Figure 2 indicates that the most occurring word within the corpus is the word "development". Table 1 shows how each top occurring words was used.

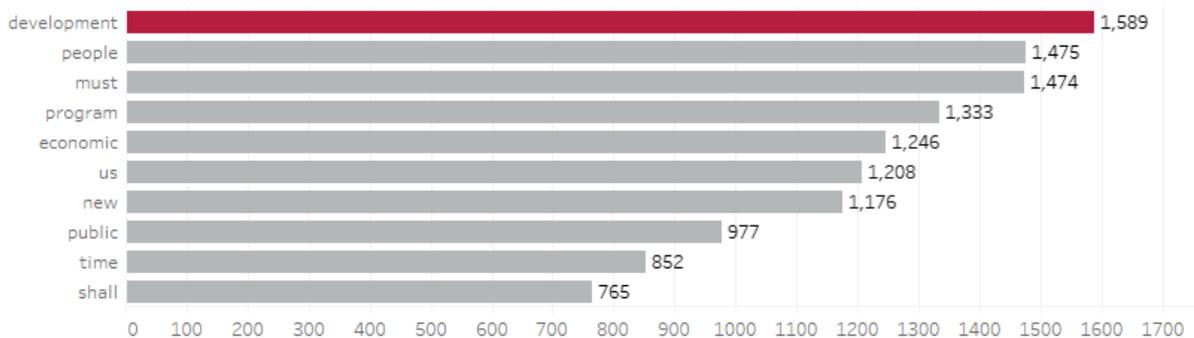

*Figure 2.* Frequency of the top 10 most occurring words found in the dataset.

Table 1. Sample sentences of the top 10 words from the dataset

| Rank | Word | Sample Sentences |
|---|---|---|
| 1 | development | "Economic growth, we have put energy development among our highest priorities." |
| 2 | People | "Every measure must be taken to protect the people." |
| 3 | Must | "We must achieve a viable consensus on an authentic family welfare program that is responsive both to the Constitutional mandate and the challenge of a growing population." |
| 4 | Program | "I responded with an economic reform program aimed at recovery in the short, and sustainable growth in the long run." |
| 5 | Economic | "The steep rise in primary energy consumption confirms the dramatic increase in economic activities." |
| 6 | Us | "Across our country, millions of Filipinos are doing that everyday: giving the sweat of their brows and the strength of their arm, above all their ingenuity and energy, to make a better future for us all." |
| 7 | New | "Fresh funding is needed to effect adjustments in the industrial structure, to assist new ventures, and to support our social programs." |
| 8 | Public | "We will intensify public health programs and pursue the various drug policies that will make essential drugs more accessible and affordable." |
| 9 | Time | "The time for planning is over; the time for action is now." |
| 10 | Shall | "Continuing progress shall be our common accountability." |

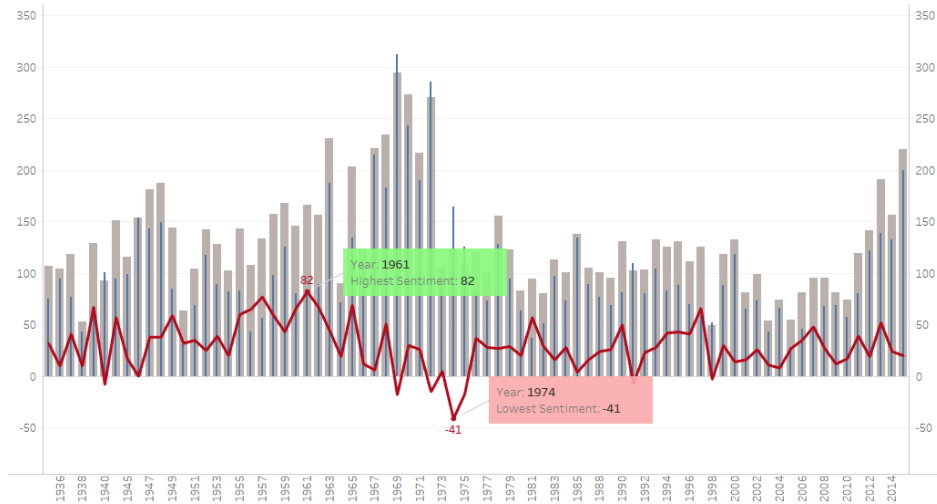

*Figure 3.* Sentiment values of SONA from 1935 to 2016.

Figure 3 indicates that it was 1961 has the highest positivity (i.e., 82) and the lowest was in 1974 (i.e., -41).

Table 2 shows a sample of positive and negative sentences using sentiment analysis.

Table 2. Sample sentence that shows Positive and Negative Sentiments during the SONA

| President | Positive | Negative |
|---|---|---|
| Quezon | "Your swift action on the defense measures I am proposing will prove the earnestness of our determination to be, and forever to remain, free and independent." | "Tactically, a fleet cannot operate as a purely defensive force and is useless unless it can proceed to see and engage its enemy beyond the limits of its own bases." |
| Osmeña | "In this most crucial hour of our history, I look forward to our destiny unafraid, confident that, God willing, ours will be a happy, progressive and prosperous land." | "I am aware that our means at the moment are inadequate." |
| Roxas | "We shall do our best to remedy the economic ailments, the best that is within our means and our under-standing." | "Our own agricultural experts tell us that we may expect a critical rice shortage in our land far exceeding anything that we have known up to this time, probably beginning in August." |
| Quirino | "This past year additional relief accrued to our people from a substantial increase in employment." | "Here the dissident elements were terrorizing the population with kidnappings and depredations." |
| Magsaysay | "Our greatest task in improving the welfare of the masses is to ensure that | "Our housing problem defies imagination." |

| | | opportunities for employment are generated." |

Table 2. Sample sentence that shows Positive and Negative Sentiments during the SONA (continuation)

| President | Positive | Negative |
|---|---|---|
| Garcia | "Matching its acquisition of a widening range of improvements and benefits, labor is developing a deeper sense of responsibility which is helpful to industrial peace." | "Unfortunately, the envisioned conditions in the accepted program of development failed to materialize within the estimated time." |
| Macapagal | "To create jobs, we will attract investments." | "This is a war we will wage on behalf, and with the rage, of all the victims: those whose businesses were ruined by extortion, those held down by poverty in fear, those whose lives were snuffed out by addiction, and those taken hostage and killed." |
| Marcos | "Let this message go forth to businessmen: Our faith in free enterprise demands that we accept the consequences of this bold adventure." | "We are faced with the irony of having to religiously invest a very high proportion of our limited resources in education and yet experiencing so much waste in high dropout rates and in the incidence of unemployment among college graduates." |
| Aquino | "That pride and that confidence rest, however, on their continuing faith in the one solid and undeniable achievement of the great moral exertion of our people: the establishment of a democratic government under an honest and dedicated leadership." | "Still, our defense expenditures are the lowest in ASEAN, and yet no country's security is so seriously threatened as ours." |
| Ramos | "Finally, I want to endorse in the strongest terms the passage of an act strengthening the Metro Manila Authority." | "But our people still live under the weight of many problems." |
| Estrada | "We will push for a tax amnesty that will give peace of mind to delinquent taxpayers and tax credits to honest ones." | "It is worse in the countryside, where 40 percent of Filipinos still work for one-third of wages in the city." |
| Arroyo | "We shall redeem in earnest the promise of land reform, a commitment that spans several presidents." | "We inherited very difficult problems." |
| Aquino III | "The mandate we received last May 10 is testament to the fact that the Filipino | "We will not allow another NBN-ZTE scandal to happen again." |

continues to hope for true change."

Table 3 presents how the sentiments in the presidents' speeches can change from their first and final speeches.

Table 3. First and Last Year Sentiments of each Presidents' Speeches

| President | Inaugural SONA | Final SONA |
|---|---|---|
| Quezon (1935 - 1941) | 14.28 | 23.40 |
| Roxas (1946 - 1948) | 23.01 | 36.67 |
| Quirino (1949 - 1953) | 39.33 | 17.24 |
| Magsaysay (1954 - 1957) | 11.47 | 31.49 |
| Garcia (1958 - 1961) | 24.89 | 32.92 |
| Macapagal (1962 - 1965) | 21.28 | 29.58 |
| Marcos (1966 - 1985) | 9.82 | 17.74 |
| Aquino (1987 – 1991) | 4.76 | 13.40 |
| Ramos (1992 – 1997) | 9.31 | 19.70 |
| Estrada (1998 – 2000) | 1.52 | 3.21 |
| Arroyo (2001 – 2009) | 11.75 | 13.86 |
| Aquino III (2010 – 2015) | 9.65 | 42.85 |

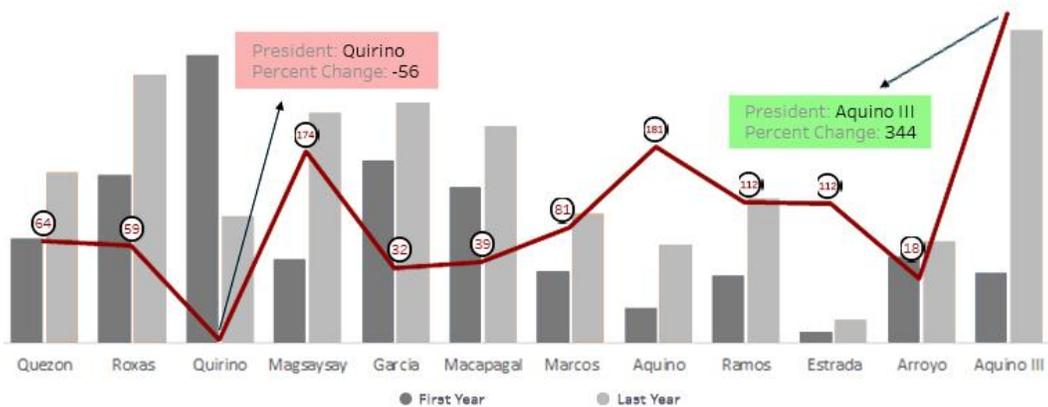

*Figure 4.* Comparison of Sentiment Values (in percent) between the First and Last Speeches of the Philippine Presidents.

Figure 4 indicates that in terms of percent changed in sentiment values between first and last SONA, it was Aquino III who have the highest positivity changed (344%) while Quirino was the only president to lower positivity when compared to his last SONA (-56%).

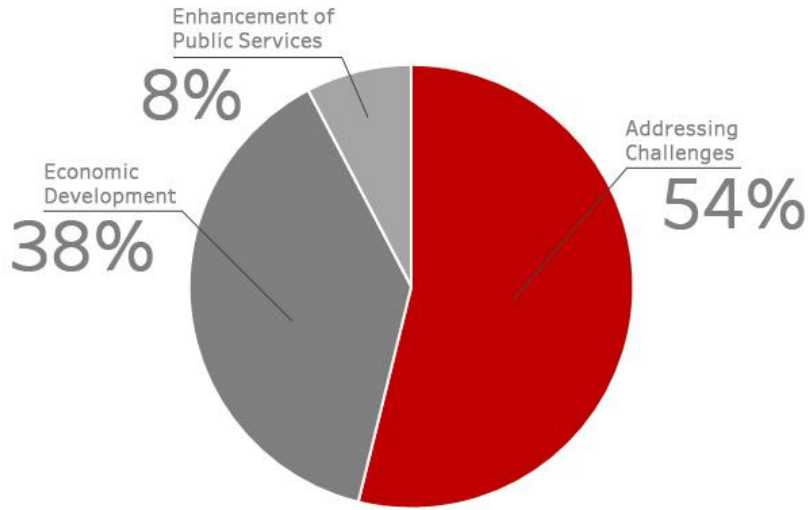

*Figure 5.* Distribution of inferred topics from presidents' speeches

Figure 5 and Table 4 shows that there are there (3) likely key topics within the dataset: economic development (38%), enhancement of public services (8%), and addressing challenges (54%).

Table 4. Inferred Topics using LDA

| Topic | Words | Sample Sentences | Label |
|---|---|---|---|
| 0 | "year", "development", "increase" "program", "economy", "land", "fund", "project", "production", "last" | "Community development shall be pursued with greater momentum and depth to develop a new sense of values and to further strengthen our social, economic, and political base in an atmosphere of mass participation and mass involvement." – Marcos<br><br>"Greater efforts should be made to diffuse the benefits and balance the economic development among the rural regions of the country." – Garcia<br><br>"In this endeavor we have addressed ourselves, first of all, to the task of rural development." – Magsaysay | Economic Development |
| 1 | "government", "make", "public", "law", "give", "need", "system", "also", "shall", "measure" | "We should give full support to the improvement of instruction and facilities in public agricultural, trade, and technical schools." – Magsaysay<br><br>"Indeed the challenge of adequacy on the basic complements of public service and physical facilities seems insurmountable; and our need for public works seems unlimited." – Marcos<br><br>"To provide employment to long unemployed people, the construction of public works projects in the troubled area was speeded up." – Quirino | Enhancement of Public Services |
| 2 | "must", "people", "nation", "country", "world", "national", "many", "problem", "today", "time" | "The welfare of the people of those areas will be safeguarded." – Roxas<br><br>"You must drop all your criminal activities because terrorism, bombings, and violence have no place in a civilized society." – Estrada<br><br>"Given our limited resources, we must improve our ratio of patrols to contacts, and our capability to maximize such combat opportunities as present themselves." - Aquino | Addressing Challenges |

Table 5 illustrates the inferred topics per president and which of these topics can be highly attributed to each president.

Table 5. LDA model score

| President | Score | Most Inferred Topic |
|---|---|---|
| Quezon | .395 | Enhancement of Public Services |
| Osmeña | .625 | Addressing Challenges |
| Roxas | .359 | Addressing Challenges |
| Quirino | .374 | Economic Development |
| Magsaysay | .384 | Economic Development |
| Garcia | .419 | Economic Development |
| Macapagal | .364 | Economic Development |
| Marcos | .368 | Economic Development |
| Aquino | .493 | Addressing Challenges |
| Ramos | .430 | Addressing Challenges |
| Estrada | .471 | Addressing Challenges |
| Arroyo | .446 | Addressing Challenges |
| Aquino III | .436 | Addressing Challenges |

## CONCLUSION AND RECOMMENDATIONS

It shows that official documents can contain meaningful information waiting to be discovered and data mining methods particularly sentiment analysis and LDA model can be used to understand these large textual data. The study was able to determine the key topics within the presidents' speeches from 1935 to 2016. These speeches are likely to focus on three main themes: economic development, enhancement of public services, and addressing challenges. The findings provided a glimpse of how Philippine presidents used their annual addresses to portray their visions for the country. The analysis further implied that from 1935 to 2016, the presidents are trying to solve and address the same perennial problems.

Future researchers may collect other official speeches made by a president during their term and subject them to the same methodology used in this study. Furthermore, the dataset may be requested from the authors for further analysis and comparison. For example, how speeches either in part or in whole reflect the preamble or foundations of the Philippine constitution.

## ACKNOWLEDGEMENT

The authors are indebted to Don Honorio Ventura State University and to the University of the East for funding this study. The dataset is available upon request to the authors of this study. The extended paper of this study is published and can be found in Cogent Social Sciences, volume 7, issue no. 1 under the Politics and International Relation section.